\documentclass[prc,showpacs,groupedaddress,floatfix,twocolumn]{revtex4}%
\usepackage{graphicx}
\usepackage{dcolumn}
\usepackage{bm}
\usepackage{amsmath}
\usepackage{amsfonts}
\usepackage{amssymb}
\usepackage{amsmath}
\usepackage{amsfonts}
\usepackage{amssymb}
\usepackage{graphicx}
\usepackage{color}%
\usepackage{hyperref}

\providecommand{\U}[1]{\protect\rule{.1in}{.1in}}

\def\symbolfootnote[#1]#2{\begingroup\def\thefootnote{\fnsymbol{footnote}}\footnote[#1]{#2}\endgroup}

\begin{document}

\title{Relativistic Continuum Quasiparticle Random Phase Approximation in Spherical Nuclei}
\author{I. Daoutidis$^{1}$\symbolfootnote[0]{1. Electronic address: idaoutid@ulb.ac.be} and P. Ring$^{2}$\symbolfootnote[0]{2. Electronic address: ring@ph.tum.de}}
\affiliation{$^{1}$ Institut d'Astronomie et d'Astrophysique, Universit\'e Libre de Bruxelles, B-1050, Belgium\\
$^{2}$ Physik-Department der Technischen Universit\"at M\"unchen, D-85748
Garching, Germany}
\date{\today}

\begin{abstract}
We have calculated the strength distributions of the dipole response in spherical nuclei, ranging all over the periodic table. The calculations
were performed within two microscopic models: the discretized quasiparticle random phase approximation (QRPA) and the quasiparticle continuum RPA,
which takes into account the coupling of the single-particle continuum in an exact way. Pairing correlations are treated with the BCS
model. In the calculations, two density functionals were used, namely the functional PC-F1 and the functional DD-PC1. Both are based on
relativistic point coupling Lagrangians. It is explicitly shown that this model is capable of reproducing the giant as well as the pygmy
dipole resonance for open-shell nuclei in a high level of quantitative agreement with the available experimental observations.
\end{abstract}

\maketitle

\section{Introduction}
The investigation of the isovector giant dipole resonance (IVGDR) is one of the fundamental problems in nuclear physics and astrophysics. These
collective resonances can be studied experimentally by photon scattering ($\gamma,\gamma^{\prime}$) or photodissociation ($\gamma,n$) processes, as
well as by means of nuclear resonance fluorescence using linearly polarized and unpolarized bremsstrahlung~\cite{LBB.74}. Theoretical investigations
are based mainly on self-consistent microscopic approaches, such as the Random Phase Approximation (RPA) or the Quasiparticle RPA (QRPA), which
is a straightforward generalization of the RPA including pairing correlations.

In solving the RPA equations one should take into account all the particle-hole pairs that contribute to the excitation. The most common
but computationally expensive way of doing this is by discretizing the basis and introducing a truncation (cut-off) parameter for the
otherwise infinite configuration space. However, despite the fact that these models, in principle, enable one to reproduce experimental mean energies
and total strengths of the giant resonances, they fail to describe their finite structure. One of the reasons is that RPA or QRPA do not provide a
mechanism that produces the escape width $\Gamma^{\uparrow}$, which gives a considerable contribution to the total width of the Giant Multipole
Resonances. In addition, the space truncation which is required by the numerical calculations leads to model deficiencies, such as deviations of the
resonance
energy or mixing of spurious states coming from symmetry braking with the physical states.

There is the alternative method of continuum RPA~\cite{SB.75} which avoids a basis truncation by treating the single-particle continuum explicitly.
As a consequence, the entire configuration space is included effectively, without the need of energy cut-offs. Despite the fact that continuum RPA
(CRPA) is proven to
be a more complete and numerically faster approach than any other RPA method available, it has been applied in the past mostly in the non-relativistic
framework~\cite{LV.68,DH.68,DD.69,SB.75,BT.75,Tsa.78,UK.89,KLL.98,HS.01,Mat.01,Mat.02,KSG.02,Bor.03,NY.05,YYM.06,OM.09,MMS.09}. Applications of
relativistic continuum RPA have been restricted to cases without pairing correlations, i.e. to doubly magic nuclei~\cite{WB.88,HP.90,SRM.89,Pie.00}.
Recently a CRPA approach based on relativistic point coupling models has been developed to study giant multipole resonances for spherical double
magic nuclei in the entire periodic table~\cite{DR.09}.

Apart from the well studied GDR, the pygmy resonance (IVPR) which appears as a soft low-lying mode in the dipole spectrum of neutron rich nuclei has
still a lot to reveal~\cite{PVK.07}. A concept that attracts most of the attention is the picture of an oscillation of the excess neutrons against the
isospin
saturated proton-neutron core. Furthermore, low-lying E1 strength in unstable nuclei is currently also discussed in the context of the astrophysical
$r$-process nucleosynthesis~\cite{Khan.02}, since they can affect the neutron capture rates, which are important quantities in the determination of
the $r$-process path. The main difficulty in the study of this mode is due to the fact that its excitation energy is very close and sometimes even
below the particle emission threshold, hindering in that way its experimental identification, specially when a photodissociation ($\gamma,n$) process
is used.

In the present paper we investigate theoretically the dipole strength distribution for several even-even nuclei all over the nuclear chart.
The calculations are performed within the framework of continuum QRPA (CQRPA) as well as of conventional QRPA, where the continuum is
discretized. Pairing correlations in open-shell nuclei are treated within the BCS model. Introducing finite occupation probabilities for
the single-particle spectrum of relativistic mean field (RMF) theory is a very successful scheme for treating $pp$-correlations in
these nuclei. In addition dynamical pairing is applied on the RPA level with an additional $pp$-term in the effective interaction. In this
way, the approach is fully self-consistent and spurious states due to restoration of symmetries, such as translation and particle number invariance
are
properly separated from the rest of the spectrum.

It has to be emphasized, however, that the BCS model faces serious problems in nuclei with large neutron excess in the neighborhood of the drip line
where the Fermi level is close to the continuum. In these cases, one has to use full Hartree Fock Bogoliubov~\cite{Go03,DNWBCD.96,DFT.84} or
Relativistic
Hartree Bogoliubov (RHB)~\cite{GEL.96,SR.02,Vretenar05} theory which treats pairing correlations in a more consistent way. But, since we concentrate
here on stable nuclei far from the drip lines, BCS is a suitable model.

The paper is organized as follows. In section \ref{CQRPA} we briefly describe the Continuum Qquasiparticle RPA based on relativistic point
coupling Lagrangians. In Section  \ref{E1}  we perform calculations on isovector dipole and pygmy resonances for several Sn isotopes and compare the
results using the various successful parameter sets. Conclusions are drawn in Sec. \ref{summary}.

\section{The point coupling model}
\label{CQRPA}
As in all relativistic models, the nucleons are described as point like Dirac particles. In contrast to the Walecka model~\cite{Wal.74}, however,
where these particles interact by the exchange of effective mesons with finite mass, point coupling models~\cite{MM.89,HMM.94} neglect mesonic degrees
of freedom and consider only interactions with zero range. In principle, these models are similar to the Nambu Jona-Lasinio model~\cite{NJL.61a} used
extensively in hadron physics. There is, however, an important difference: in order to obtain a satisfactory description of the nuclear surface
properties one also needs gradient terms in the Lagrangian simulating a finite range of the interaction. In this work we use two different point
coupling Lagrangians; the set PC-F1 introduced by Buervenich et al. in Ref.~\cite{BMM.02} and the more recent set DD-PC1 by Nik\ifmmode \check{s}\else
\v{s}\fi{}i\ifmmode \acute{c}\else \'{c} et al.~\cite{NVR.08}. In both cases, the Lagrangian is represented in terms of the nucleon scalar~(S),
vector~(V)
and isovector-vector~(TV) fields:
\begin{eqnarray}
\label{Lagrangian}
 \mathcal{L}&=&\bar{\psi}(i\gamma\cdot\partial-m)\psi \nonumber \\
&-&\sum_{i}\frac{1}{2}\alpha_{i}[\rho](\bar{\psi}\hat{\Gamma}_i\psi)(\bar{\psi}\hat{\Gamma}_i\psi)
-\frac{1}{2}\delta_{i}(\partial_{\nu}\bar{\psi}\hat{\Gamma}_i\psi)(\partial^{\nu}\bar{\psi}\hat{\Gamma}_i\psi)\nonumber\\
&-& e^2(\bar{\psi}\hat{\Gamma}_{C}\psi)(\bar{\psi}\hat{\Gamma}_{C}\psi),
\end{eqnarray}
where the Dirac vertices $\hat{\Gamma}_i~(i=S,V,TV)$ and the electromagnetic vertex $\hat{\Gamma}_C$ have the explicit form:
\begin{equation}
\hat{\Gamma}_{S} = 1,~~\hat{\Gamma}_{V} = \gamma_{\mu},~~\hat{\Gamma}_{TV} = \gamma_{\mu}\vec{\tau},~~
\hat{\Gamma}_{C} = \gamma_\mu\frac{(1-\tau_{3})}{2}.
\end{equation}
The coupling constants $\alpha_{i}$ depend on the density and this density dependence is different for the two forces.
While, for PC-F1 each $\alpha_{i}[\rho_i]$ in the various spin-isospin channels depends on the corresponding local densities $\rho_i$,
for DD-PC1 all the couplings $\alpha_{i}[\rho]$ depend on the baryon density $\rho=\rho_V$ alone. In particular, one
has:
\begin{equation}
\alpha_{i}[\rho]=\left\{\begin{array}{ccc}a_{i}+b_{i}\rho_{i}+c_{i}\rho_{i}^{2}~~~~~~& {\rm for} & {\rm PC-F1},~~ \\
a_{i}+(b_{i}+c_{i}x)e^{-d_{i}x}& {\rm for}& {\rm DD-PC1},\end{array}\right.
\end{equation}
where $x=\rho/\rho_{sat}$ denotes the nucleon density in units of the saturation density in symmetric nuclear matter.

In addition, DD-PC1 unlike its predecessors has not been adjusted to spherical nuclei but to ab-initio calculations together 
with nuclear matter data and to a large set of axially deformed nuclei. The two sets are listed in
Table~\ref{tab1} and they have been tested in the calculation of many ground-state properties of spherical and deformed nuclei all over the periodic
table~\cite{NVR.05}. The results are very well comparable with reasonable effective meson-exchange interactions~\cite{LKF.09,DD-ME2,DL.05}.

However, the success of a particular interaction relies on the ability to reproduce, apart from the static properties of a broad range of nuclei, also
their dynamical properties, as for instance the properties of collective multipole resonances via the microscopic RPA approaches, which we
will discuss in the next section.
\begin{table}[t]
\centering
\renewcommand{\arraystretch}{1.5}%
\begin{tabular}[c]{lr@{.}l r@{.}l}
\hline\hline
~~~& \multicolumn{2}{c}{PC-F1}&\multicolumn{2}{c}{DD-PC1}\\
\hline
~~~~$a_{S}$& -14 & 935894~~[fm$^{-2}$]&-10 & 0462~~[fm$^{-2}$]\\
~~~~$b_{S}$& 22 & 994736~~[fm$^{-5}$]&-9 & 1504~~[fm$^{-2}$]\\
~~~~$c_{S}$& -66 & 769116~~[fm$^{-8}$]&-6 & 4273~~[fm$^{-2}$]\\
~~~~$d_{S}$&\multicolumn{2}{c}{~~}&1& 3724\\
~~~~$\delta_{S}$& -0 & 634576~~[fm$^{-2}$]&-0& 8149~~[fm$^{-4}$]\\\hline
~~~~$a_{V}$ & 10 & 098025~~[fm$^{-2}$]&5 & 9195~~[fm$^{-2}$]\\
~~~~$b_{V}$ & 0 & 0&8 & 8637~~[fm$^{-2}$]\\
~~~~$c_{V}$ &-8 & 917323~~[fm$^{-8}$]&0 & 0\\
~~~~$d_{V}$ &\multicolumn{2}{c}{~~} &0& 6584\\
~~~~$\delta_{V}$ & -0 & 180746~~[fm$^{-2}$]&0& 0\\\hline
~~~~$a_{TV}$ & 1 & 350268~~[fm$^{-2}$]&0 & 0\\
~~~~$b_{TV}$ & 0 & 0&1 & 8360~~[fm$^{-2}$]\\\
~~~~$c_{TV}$ &0 & 0&0 & 0\\
~~~~$d_{TV}$ &\multicolumn{2}{c}{~~}&0 &6403\\
~~~~$\delta_{TV}$ & -0 & 063680~~[fm$^{-2}$]&0 & 0\\\hline\hline
\end{tabular}
\caption{The coupling constants for the density functionals PC-F1 and DD-PC1 resulting from the fitting procedure in Ref.~\cite{BMM.02} and
Ref.~\cite{NVR.08} respectively. The parameters $d_{i}$ of DD-PC1 are dimensionless.}
\label{tab1}%
\end{table}

\subsection*{Linear response theory}

If a nucleus is exposed to an external field, such as in the photoabsorption process, the strength function
\begin{equation}
\label{strenghtfunction}
S(\omega)=-\frac{1}{\pi}Im\sum_{\alpha\beta} F_{\alpha}^{\ast}R^{}_{\alpha\beta}(\omega)F^{}_{\beta}
\end{equation}
measures the change on the nuclear density due to the influence of this field. The Greek indices $\alpha,\beta$  indicate the various degrees  of
freedom of a nucleus ($r,L,S,T$) while $F$ is the operator of the external field.

If we consider only small amplitude variations of the density taking into account only one particle-one hole ($1ph$) excitations, the response
function
$R(\omega)$ can be deduced from the \textit{linearized Bethe-Salpeter} equation:
\begin{equation}
\label{Bethe-Salpeter}
R^{}_{\alpha\beta} (\omega)=R_{\alpha\beta}^{\,0}(\omega)+\sum_{\gamma\delta}\,\,R^0_{\alpha\gamma}(\omega)V^{ph}_{\gamma\delta}R^{}_{\delta\beta}(\omega).
\end{equation}

This method is usually referred to as a response function formalism of the Random Phase Approximation, to distinguish from the more frequently used
configuration space formalism~\cite{RMG.01}. The reason we chose this forma\-lism is because it is essential for an exact treatment of the continuum
coupling.

\subsubsection*{The residual interaction}
As explained in textbooks~\cite{RS.80,MUN.06}, the residual interaction $V^{ph}_{\alpha\beta}$ of Eq.~(\ref{Bethe-Salpeter}) is connected to the
static problem via the second derivative of the energy functional
\begin{equation}
\label{energy_variation}
V_{\alpha\beta}^{ph}= \frac{\delta^{2}E[\hat{\rho}]}{\delta\hat{\rho}_{\alpha}\delta\hat{\rho}_{\beta}}
\end{equation}
and hence, it depends on the same coupling constants $\alpha_{i}[\hat{\rho}]$ and their derivatives with respect to the densities. In this way, one
ensures a fully
self-consistent treatment of the dynamical problem. The point coupling scheme allows one to write $V^{ph}$ as a sum of separable terms, simplifying
considerably the numerical solution of the Bethe-Salpeter equation. In particular, one can write:
\begin{equation}
\label{V_separable}
V^{ph}(1,2)=\sum\limits_{c}\int\limits_{0}^{\infty}dr~Q_{c}^{(1)}(r)~\upsilon^{}_{cc^{\prime}}(r)~Q_{c^{\prime}}^{\dag(2)}(r),
\end{equation}
where the upper indices (1) and (2) indicate that  these operators act on the "coordinates" $1=(r_{1}\Omega_{1}s_{1}d_{1}t_{1})$ and
$2=(r_{2}\Omega_{2}s_{2}d_{2}t_{2})$.
\begin{table}[t]
\centering\renewcommand{\arraystretch}{1.5}
\begin{tabular}
[c]{|c|ccc|}\hline
PC-F1& S & V & TV \\\hline
S & $F_{S}[\rho_{S}]$ &           0       &            0         \\
V &         0         & $F_{V}[\rho_{V}]$ &            0         \\
TV&         0         &           0       &  $F_{TV}[\rho_{TV}]$ \\\hline
\end{tabular}
\caption{The structure of the matrix $\upsilon_{cc^{\prime}}$ in spin-isospin space for the PC-F1 parametrization. The functional
$F_{i}[\rho]=\alpha_{i}[\rho_{i}]+2\alpha_{i}'[\rho_{i}]\rho_{i}+1/2\alpha_{i}''[\rho_{i}]\rho_{i}^{2}+\delta_{i}\Delta$.}
\label{tab2}
\end{table}

Each separable term is characterized by ($c,r$), the channel index $c$ given by the discrete numbers $\{D,S,L,J,T\}$ and the radial mesh point $r$.
In point coupling models under investigation, there are
overall seven channels; one scalar S, three isoscalar V and three isovector TV vectors. Assuming a coordinate mesh of 50 points, the final size of
the interaction matrix is not larger than $350\times\,350$. The corresponding channel vertices $\hat{Q}_{c}^{(1)}(r)$ are local single-particle
operators
\begin{equation}
\label{channel-operator}
\hat{Q}_{c}^{(1)}(r)=(-)^{S_{c}}\frac{\delta(r-r_1)}{rr_1} \hat{\Gamma}^{(1)}_{c} Y_L (\Omega_{1}).
\end{equation}
Therefore, the Bethe-Salpeter equation (\ref{Bethe-Salpeter}) becomes:
\begin{equation}
\label{reduced-bs}
R ^{}_{cc^{\prime}}(\omega)=R_{cc^{\prime}}^{\,0}(\omega)+
\sum_{c_{1}c_{2}}R_{cc_{1}}^{\,0}(\omega)\upsilon^{}_{c_{1}c_{2}}R^{}_{c_{2}c^{\prime}}(\omega),
\end{equation}
which has the same formal solution as  in Eq.~(\ref{Bethe-Salpeter}). For the continuous variables of the channel index $c$, such as the radial
coordinate $r$, Eq.~(\ref{reduced-bs}) is an integral equation.

The interaction $\upsilon_{cc^{\prime}}$  can be expressed as a matrix in the spin-isospin space, indicated by the index $c$. Apart from the direct
terms $\alpha_{i}[{\hat\rho}]$ of the Lagrangian, $\upsilon_{cc^{\prime}}$ also consists of the so called rearrangement terms, as a result of the
double variation of Eq.~(\ref{energy_variation}). Furthermore, the gradient term is described by the operator~\cite{DR.09}:
\begin{equation}
\label{gradient}
\Delta=r^{2}\overleftarrow{\partial}_{r}\frac{1}{r^{2}}\overrightarrow
{\partial}_{r}+\frac{L(L+1)-2}{r^{2}}.
\end{equation}
In Table~\ref{tab2} and ~\ref{tab3}, the exact form of the interaction matrices is displayed for
the two parameter sets PC-F1 and DD-PC1. Further details
can be found also in Refs.~\cite{DR.09,Dao.09}.

\begin{table}[t]
\centering\renewcommand{\arraystretch}{1.5}
\begin{tabular}
[c]{|c|ccc|}\hline
DD-PC1&               S                     &                V              &               TV              \\\hline
S     & $\alpha_{S}[\rho]+\delta_{S}\Delta$ & $\alpha_{S}'[\rho]\rho_{S}$   &                0              \\
V     & $\alpha_{S}'[\rho]\rho_{S}$         & $ F_{V}[\rho]$                & $\alpha_{TV}'[\rho]\rho_{TV}$ \\
TV    &               0                     & $\alpha_{TV}'[\rho]\rho_{TV}$ &  $\alpha_{TV}[\rho]$          \\\hline
\end{tabular}
\caption{The structure of the matrix $\upsilon_{cc^{\prime}}$ for the DD-PC1 parametrization. The functional
$F_{V}[\rho]=(\alpha_{V}[\rho]+2\alpha_{V}'[\rho]\rho+1/2\alpha_{V}''[\rho]\rho^{2}) +1/2\alpha_{S}''[\rho]\rho_{S}^{2} +
1/2\alpha_{TV}''\rho_{TV}^{2}$.}
\label{tab3}
\end{table}
\begin{table}[b]
\centering\renewcommand{\arraystretch}{1.5}
\begin{tabular}
[c]{|c|ccc|}\hline
  &            S               &            V              &          TV                 \\\hline
S &            0               &            0              &            0                \\
V &            0               & $\frac{1}{4}\upsilon_{C}$ & $-\frac{1}{4}\upsilon_{C}$  \\
TV&            0               &$-\frac{1}{4}\upsilon_{C}$ & $\frac{1}{4}\upsilon_{C}$   \\\hline
\end{tabular}
\caption{The structure of the channel matrix $\upsilon^{C}_{cc^{\prime}}$ for the Coulomb interaction.}
\label{tab35}
\end{table}
Apart from the Coulomb force
\begin{equation}
\upsilon_{\text{C}}(r,r^{\prime})=\frac{4\pi e^2}{2L+1}\cdot\frac{r_{<}^{L}}{r_{>}^{L+1}}
\end{equation}
all other interaction terms are diagonal in $r$. Here, $r_{<}$ and $r_{>}$ are the smaller and the greater of $r$ and $r^{\prime}$.
This force breaks isospin symmetry. Hence, one has to expand
$\upsilon_{\text{C}}$ in its spin-isospin components:
\begin{eqnarray}
\label{vcoulomb}
V_{C}(1,2)&=&(\frac{1}{2}(1+\tau_{3}))^{(1)}\frac{e^2}
{|\mathbf{r}_{1}\mathbf{-r}_{2}|}(\frac{1}{2}(1+\tau_{3}))^{(2)}  \\
&=&\frac{1}{4}\frac{e^2}{|\mathbf{r}_{1}\mathbf{-r}_{2}|}\nonumber\\
&\times&(\mathbf{1}^{(1)}\mathbf{1}^{(2)}-\mathbf{1}^{(1)}\tau_{3}^{(2)}-
\tau_{3}^{(1)}\mathbf{1}^{(2)}+\tau_{3}^{(1)}\tau_{3}^{(2)}).\nonumber
\end{eqnarray}
This means that in spin-isospin space, the Coulomb interaction can be split into four parts, but thankfully not into four additional channels. In
particular, the two diagonal terms are added to the already existing isoscalar and isovector vector channels, while the other two correspond to
off-diagonal
terms. This leads to a Coulomb matrix $\upsilon^{C}_{cc^{\prime}}(r,r^{\prime})$ as shown in Table~\ref{tab35}.

\subsubsection*{The free response function}
The free response function $R^{0}_{cc^{\prime}}(\omega)$ is the key point for a successful description of the RPA problem, since it includes all the
microscopic properties of the nucleus under investigation. In the discrete quasiparticle space, i.e. in the spectral representation, it can be written as:
\begin{eqnarray}
\label{R0_2qp}
\mathcal{R}_{\text{2qp}}^{0} &=&\sum_{\alpha\leq \beta}^{E_{ph}^{max}}\frac{1}{1+\delta_{\alpha\beta}}\eta_{\alpha\beta}
^{S}\langle \alpha||Q^{\dag}_{c}||\beta\rangle_{r}\eta_{\alpha\beta}^{S^{\prime}}\langle \alpha||Q_{c^{\prime}}||\beta\rangle_{r^{\prime}}
\nonumber\\
&&\times\left(\frac{1}{\omega-E_{\alpha\beta}+i\eta}-\frac{1}{\omega +E_{\alpha\beta}+i\eta}\right),
\end{eqnarray}
where the indices $\alpha$ and $\beta$ refer to the two-quasiparticle states in the excitation, while $|\alpha\rangle_r=|\alpha(r)\rangle$ and
$|\beta(r)\rangle$ are the corresponding radial single-particle wave functions. The factors $\eta_{\alpha\beta}^{c}=u_{\alpha}v_{\beta}+
(-)^{S_{c}}u_{\alpha}v_{\beta}$ include the BCS occupations $u_{\alpha}$ and $v_{\alpha}$. The energy $E_{\alpha\beta}=E_{\alpha}+E_{\beta}$ is the
sum of two quasiparticle energies. In addition, the smearing parameter $\eta$ is used in Eq.~(\ref{R0_2qp}) in order to avoid a divergence of the
free response function and its value is often adjusted to the observed width of the giant resonances. It has to be emphasized that according to the
no-sea approximation in relativistic RPA the sum in Eq.~(\ref{R0_2qp}) includes also transitions to negative energy solutions~\cite{DF.90,RMG.01} 
since the vacuum polarization is neglected. 

One of the important questions which arises by using the spectral representation of Eq.~(\ref{R0_2qp}) is the question of completeness of
the basis. Numerical limitations require a basis truncation, which can play a role on the final solution. In practical calculations, an energy
cut-off $E^{ph}_{cut}=500~$MeV for the particles and a corresponding $E^{ah}_{cut}=-2000~$MeV for the antiparticles is required to achieve a
converged solution. This fact, combined with the additional excitations to the anti-particles states leads to a very large configuration
space~\cite{RMG.01}.

In continuum QRPA however, the situation is very different. Here, one makes use of the non-spectral representation of the free response
function, which is given by:
\begin{eqnarray}
\label{R0_cont}
\mathcal{R}^{0}_{cc^{\prime}} = \sum_{\alpha}
&u_{\alpha}^{2}&
\langle\alpha(r)|\hat{Q}^{\dag}_{c}g^{}_{\kappa_\alpha}(\omega+\varepsilon_{\alpha})\hat{Q}_{c^{\prime}}|\alpha(r^{\prime})\rangle    \\
&+&
\langle\alpha(r)|\hat{Q}^{\dag}_{c^{\prime}}g^{}_{{\kappa_\alpha}}(-\omega+\varepsilon_{\alpha})\hat{Q}_{c}|\alpha(r^{\prime})|\rangle,  \nonumber
\end{eqnarray}
where the relativistic Green's function is:
\begin{eqnarray}
\label{greens-continuum}
g^{}_{\kappa}(E)=
\left\{\begin{array}[c]{cc}|w_{\kappa}(r)\rangle\langle u^\ast_{\kappa}(r^{\prime})|/W &
\text{~~for}
\,\,r>r^{\prime}\\
|u_{\kappa}(r)\rangle\langle w^\ast_{\kappa}(r^{\prime})|/W & \text{~~for}\,\,r<r^{\prime}
\end{array}
\right..
\end{eqnarray}
The two-dimensional spinors $|u_{\kappa}(r)\rangle={f_{u}(r) \choose ig_{u}(r)}$ and $|w_{\kappa}(r)\rangle={f_{w}(r) \choose ig_{w}(r)}$ are the
regular and the irregular wavefunctions of a continuum state with the quantum number $\kappa=(lj)$ and the energy $E$. Further details are given in Ref.~\cite{DR.09}.
Since $|w_{\kappa}(r)\rangle$ behaves as an outgoing wave at $r\rightarrow\infty$, it is a complex quantity. In this way, the strength function in 
Eq.~(\ref{strenghtfunction}) has a non-vanishing imaginary part without the need of any smearing parameter, as in Eq.~(\ref{R0_2qp}). 
The denominator W describes the Wronskian of the system:
\begin{equation}
W=\langle\,w_\kappa(r)|u_{\kappa}^{*}(r)\rangle
= f_{w}(r)g_{u}(r)-g_{w}(r)f_{u}(r)
\end{equation}
and is independent of $r$. In the particular case where the energy of the continuum state meets the value of an eigenstate
$E=\omega+\varepsilon_{\alpha}=\varepsilon_{p}$, the two wavefunctions $|u_{\kappa}\rangle$ and $|w_{\kappa}\rangle$ become identical and the
Green's  function diverges, similar to the case in Eq.~(\ref{R0_2qp}). So, the two expressions of the free response functions
in Eqs.~(\ref{R0_2qp}) and (\ref{R0_cont}) are in principle identical, but in practice and due to the space truncation of the spectral
representation, differences are expected.

Consequently, one of the big advantages of the non-spectral or continuum representation Eq.~(\ref{R0_cont}) is the inclusion of the entire
configuration space, without the need of any truncation technique. This is achieved by summing over blocks of well defined quantum numbers $\kappa$.
The wavefunctions are exact and obey the proper boundaries at infinity. In this way, the numerical effort can be reduced to more than one
order of magnitude, as compared to the conventional RRPA approaches, where the continuum is discretized~\cite{NVR.05}. Finally, as far as the
solution is concerned, the inclusion of the full space ensures a more realistic description of the nuclear collective properties and better
agreement with the experimental data~\cite{DR.09}.

\begin{figure*}[!t]
\centering
\includegraphics[width=300pt]{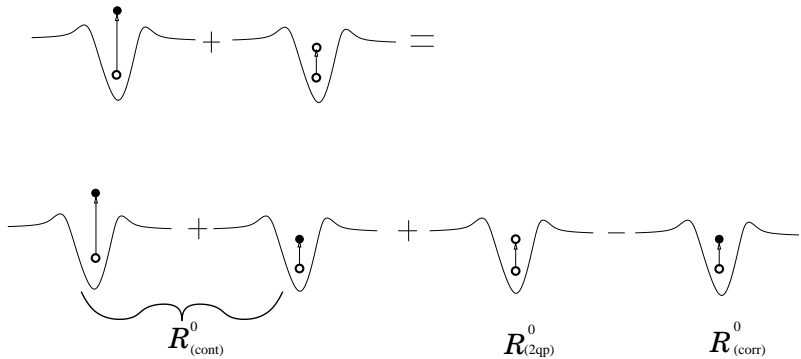} \newline
\caption{Various configurations used for the calculation of the free quasiparticle response. Filled circles ($\bullet$) refer to a pure particle
($v^{2}_{k}=0$), while empty circles ($\circ$) indicates the quasiparticles ($v^{2}_{k}>0$). Details are given in the text.}
\label{fig01}
\end{figure*}

Despite its simplicity, a problem arises when pairing correlations are included. From Eq.~(\ref{R0_cont}), we see that all levels lying in the
continuum have $u_{\beta}=1$, i.e. they are all considered as pure particles. But this is not the case in the QRPA model, where levels close to the
Fermi surface (some of them above the single-particle continuum limit) must be treated as quasiparticles. That implies that in order to be
consistent,
we need to modify the continuum representation to properly account for the states within the space, where pairing is active.

In Fig.~\ref{fig01}, we schematically illustrate our technique, which has been used in the past to define the continuum QRPA in non-relativistic
models~\cite{KLL.98,HS.01}. In this figure, the allowed transitions are displayed as a sum of three different terms. At first, the continuum term
$\mathcal{R}_{cont}^{0}$ of Eq.~(\ref{R0_cont}) includes the transitions to the entire single-particle space.

The two-quasiparticle excitations within the pairing active space are treated by the two-quasiparticle response function Eq.~(\ref{R0_2qp}) up
to the pairing cut-off, which does not exceed the $E_{p}=+20~$MeV.

However, in order to avoid double counting of the excitations in the active pairing space, one needs to subtract a correction term, which describes
transitions from a quasiparticle to a pure particle using the expression:
\begin{eqnarray}
\label{R0_corr}
\mathcal{R}_{\text{corr}}^{0} &=&\sum_{\alpha\leq \beta}^{E_{p}}\frac{1}{1+\delta_{\alpha\beta}}\langle
\alpha||Q_{c}^{\dag}||\beta\rangle_{r}\langle \alpha||Q_{c^{\prime}}||\beta
\rangle_{r^{\prime}}  \nonumber \\
&& \times\left\{\frac{v_{\alpha}^{2}}{\omega-\Omega_{\alpha,\beta}+i\eta}-\frac{v_{\alpha}^{2}}{\omega
+\Omega_{\alpha,\beta}+i\eta}\right.\\
&& \left.+\frac{v_{\beta}^{2}}
{\omega-\Omega_{\beta,\alpha}+i\eta}-\frac{v_{\beta
}^{2}}{\omega+\Omega_{\beta,\alpha}+i\eta}\right\},\nonumber
\end{eqnarray}
where $\Omega_{\alpha,\beta}=E_{\beta}+\varepsilon_{\alpha}-\lambda$. The indices $\alpha$ and $\beta$ run only over the partially occupied states
below the continuum limit ($\varepsilon_{\alpha}<0$). Consequently, a proper treatment of the quasiparticle CRPA requires that the response function
\begin{eqnarray}
\label{sum_term}
 \mathcal{R}^{0}(r,r^{\prime}\,\omega) =\mathcal{R}_{\text{cont}}^{0}(r,r^{\prime}\,\omega) &+&
\mathcal{R}_{\text{2qp}}^{0}(r,r^{\prime}\,\omega)\nonumber \\
 &-&\mathcal{R}_{\text{corr}}^{0}(r,r^{\prime}\,\omega)
\end{eqnarray}
is used in the Eq.~(\ref{Bethe-Salpeter}).

It has to be emphasized that all the particle states above the pairing window as well as the states in the Dirac sea do not participate in the above
calculation, since they are completely taken into account in the continuum part $\mathcal{R}_{\text{cont}}^{0}(r,r^{\prime}\,\omega)$ in an exact way.

\begin{figure*}[ht!]
\centering
\includegraphics[width=300pt]{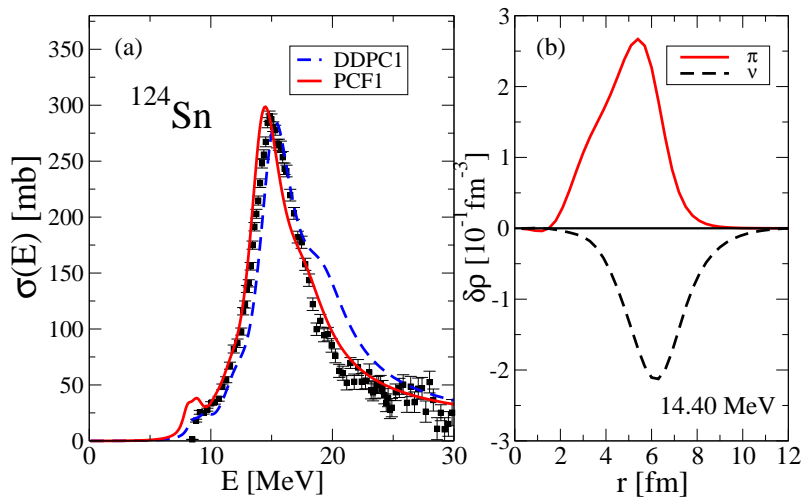}
\newline\caption{(Color online) Panel (a): The IVGDR strength distributions for $^{124}$Sn
using the forces PC-F1 (red solid line) and DD-PC1 (blue dashed line). The experimental results are taken from Ref.~\cite{FBC.69}.
Panel (b): Neutron and proton transition densities for the GDR at $E=14.40$ MeV.}
\label{fig02}
\end{figure*}
\section{Results and discussions}
\label{E1}

In the following we show the IVGDR results for various spherical open-shell nuclei, using both the continuum and the discrete QRPA models.

In a first step, the ground state of the nucleus is determined by solving the self-consistent RMF equations for the parameter sets PC-F1 and DD-PC1
given in Table~\ref{tab1}. Pairing correlations are treated within the BCS model with constant gap. For the pairing gap, either the empirical
expression $\Delta_{n,p}=12.0/\sqrt{A}$ or the gaps deduced from nuclear mass tables~\cite{CMP.10} can be used, with the difference to be
insignificant to the final outcome.

Using the single-particle wave functions and the corresponding quasiparticle energies of this static solution, we determine the free response
$\mathcal{R}^{0}$ of Eq.~(\ref{sum_term}). Finally, we solve the Bethe-Salpeter equation (\ref{reduced-bs}) to get the strength distribution
$S(\omega)$.

At the same time, we perform similar calculations using the conventional RPA approach, where the continuum is fully discretized. For those
calculations, we have used an energy cut-off $|\epsilon_{p}-\epsilon_{h}| < E^{ph}_{cut}= 300$ MeV for the configurations with particles above the
Fermi sea and $|\epsilon_{a} -\epsilon_{h}|<E^{ah}_{cut}=2000$ MeV for configurations with anti-particles in the Dirac sea.

In the left panel of Fig.~\ref{fig02}, we show the cross section for the IVGDR in the open shell nucleus $^{124}$Sn for the two parameter sets PC-F1
(red solid) and
DD-PC1 (blue dashed), as well as the experimental values from Ref.~\cite{FBC.69}. The cross section is linear to the E1 strength of
Eq.~(\ref{strenghtfunction}) according to:
\begin{eqnarray}
 \sigma(E) &=& \frac{16\pi^{3}e^{2}}{9\hbar\,c}E\,S(E)\,\, [{\rm fm}^{2}] \nonumber \\
&=& 4.022\,E\,S(E)\,\, [{\rm mb}].
\end{eqnarray}
The two parameterizations PC-F1 (red solid) and DD-PC1 (blue dashed) perform very well in reproducing the Giant Dipole resonance, as compared
to experimental data~\cite{FBC.69}. The energy-weighted sum rule for the two forces are found at $m_{1}=2043.47$ and $m_{1}=2151.85$
~$[\rm mb\cdot MeV]$ respectively. These results are in agreement with the discrete QRPA and, as usual, somewhat ($13.7\%$ and $19.7\%$) larger than
the classical Thomas-Reiche-Kuhn (TRK) sum rule:
\begin{equation}
\label{TRK}
m_{\rm TRK}= 60.0\frac{NZ}{A}=1782.32~ [\rm mb\,MeV].
\end{equation}

Due to the exact coupling to the continuum, the escape width $\Gamma^{\uparrow}$ of the IVGDR is automatically taken into account. However, in heavy
nuclei,
$\Gamma^{\uparrow}$ is very small, due to the high Coulomb and centrifugal barriers, which prevent the excited nucleon from escaping. The rest of the
total width comes mainly from the coupling to more complex configurations (2ph, 3ph, etc.). Here, this part is treated approximately by an
additional smearing parameter, which can be energy or temperature dependent~\cite{BKC.92}, according to the expression:
\begin{equation}
\Gamma(E)=\Gamma_{0}\frac{E^{2}+4\pi^{2}T^{2}}{E_{GDR}^{2}}.
\end{equation}
In our study, we have calculated the width by using $\Gamma_{0}=1.4~$MeV and $T=0$. On the right panel of Fig.~\ref{fig02} we give the proton and
neutron transition densities $\delta\rho(r)$ associated with the GDR peak at $E=14.40$~MeV. We see that the excitation at this energy has a pure
isovector character, since the neutrons are coherently oscillating against the protons.

\begin{figure}[b]
\centering
\includegraphics[width=240pt]{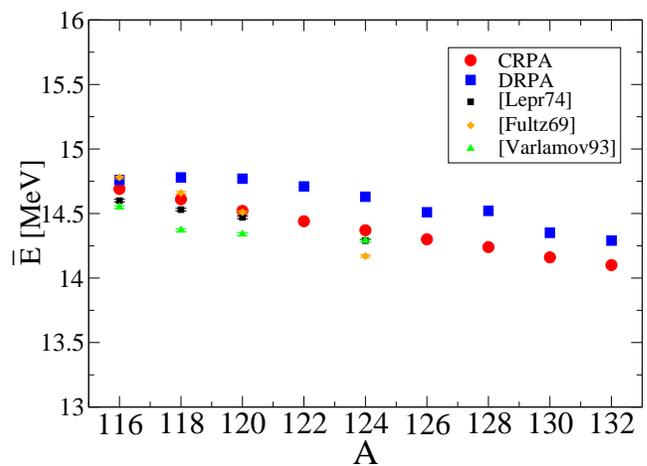}\newline\caption{(Color online) Centroid energies for several Sn isotopes using the DRPA
(blue boxes) and the CRPA (red squares) approaches. The experimental results are deduced from Refs.~\cite{LBB.74,FBC.69,Var.93}.}
\label{fig03}%
\end{figure}

\begin{table}[!t]
  \centering
\renewcommand{\arraystretch}{1.5}%
  \begin{tabular}{lrcccc}
  \hline\hline
  &  & ~~DD-PC1~~ & ~~PC-F1~~ &  ~Exp.~[MeV]& \\
  \hline
  $^{70}$Zn  & $E_{0}$   &17.50 & 16.70 & 17.25 $\pm$ 0.08 &\cite{GZ.82}\\
             & $\bar{E}$ &16.00 & 15.86 & 15.68 $\pm$ 0.02 &\\\hline
  $^{94}$Zr  & $E_{0}$   &16.60 & 15.60 & 16.67 $\pm$ 0.07 &\cite{BCH.67}\\
             & $\bar{E}$ &15.90 & 15.58 & 16.00 $\pm$ 0.01 &\\\hline
  $^{124}$Sn & $E_{0}$   & 15.40 & 14.40 & 14.67 $\pm$ 0.08& \cite{LBB.74}\\
             & $\bar{E}$ & 14.99 & 14.70 & 14.34 $\pm$ 0.02& \\\hline
  $^{130}$Te & $E_{0}$   & 15.30 & 14.60 & 14.53 $\pm$ 0.13& \cite{LBB.76}\\
             & $\bar{E}$ & 14.96 & 14.66 & 14.27 $\pm$ 0.01& \\\hline
  $^{138}$Ba & $E_{0}$   & 15.20 &14.40 & 15.29 $\pm$ 0.15 &\cite{BFC.70}\\
             & $\bar{E}$ & 14.89 &14.55 & 14.64 $\pm$ 0.01 &\\\hline
  $^{144}$Sm & $E_{0}$   & 15.10 & 14.50 & 15.37 $\pm$ 0.13& \cite{CBB.74}\\
             & $\bar{E}$ & 15.39 & 14.58 & 14.77 $\pm$ 0.02& \\\hline
  $^{208}$Pb & $E_{0}$   & 13.60 & 12.80 & 13.50 $\pm$ 0.19& \cite{VBB.70}\\
             & $\bar{E}$ & 14.13 & 13.78 & 13.52 $\pm$ 0.04& \\\hline\hline
\end{tabular}
\caption{Excitation Energy of the isovector dipole resonance for several nuclei, using both DD-PC1 and PC-F1 self-consistent interactions. The
centroid energies have been calculated in the area between $11-18~$MeV.}
\label{tab4}
\end{table}

In Fig.~\ref{fig03} the centroid energies of the IVGDR are revealed for several Sn isotopes, which have been well identified experimentally
~\cite{FBC.69,LBB.74,Var.93}. The centroid energy $\bar{E}=m_{1}/m_{0}$ is calculated in the same energy window as the one used in the
experimental analysis, i.e. between 13-18~MeV. It can be clearly seen that both discrete and continuum QRPA approaches predict a similar decrease of
the centroid energies, with respect to the mass number of the isotope, but the agreement with the experimental data is better reproduced with the
continuum approach.

In Table~\ref{tab4}, results for the two density functionals PC-F1 and DD-PC1 are compared using the continuum approach. We have
studied a series of stable spherical nuclei, which have experimentally well defined collective properties. It appears that the force PC-F1 is
in general slightly better in performance, although both sets are very successful in predicting the collective properties.

As it is for instance clearly seen in Fig.\ref{fig02}, in the case of neutron rich nuclei, in addition to the giant dipole resonance a small peak
appears at the energy region of the neutron emission threshold around $E\sim8$~MeV. This pygmy mode is classically interpreted as an oscillation of
the neutron skin against the isospin saturated proton-neutron core and it seems to have a collective character as well.

The collectiveness of this pygmy mode can be studied by plotting the evolution of its strength with respect to the mass number. As we show in the
upper panel of Fig.~\ref{fig04} the contribution of the pygmy dipole resonance (PDR) strength to the Thomas-Reiche-Kuhn sum rule increases
considerably, as we move to heavier isotopes, meaning that all the additional neutrons participate in this resonance. In particular we find that there
is a sharper increase after the double-magic isotope $^{132}$Sn, which is observed in both the continuum and the discrete approach. This can be
connected to the sharp decrease of the particle emission threshold at exactly the same isotope, as we see in the lower panel of Fig.~\ref{fig04}.

\begin{figure}[!t]
\centering
\includegraphics[width=240pt]{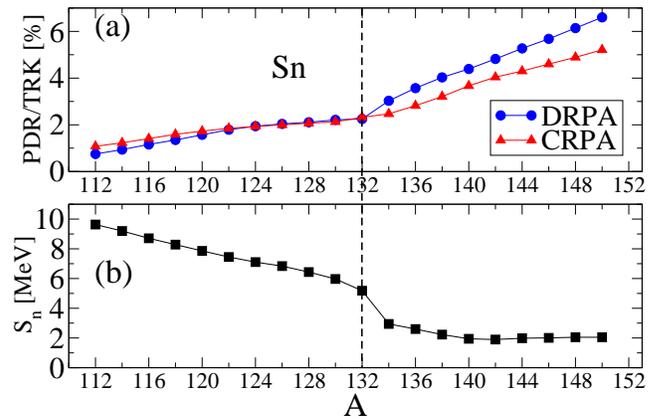}\newline\caption{(Color online) Panel (a): The ratio of the PDR strength to the TRK sum
rule for various Sn isotopes. The blue circles indicate discrete RPA results, while the red triangles are Continuum RPA results. The calculated
PDR is integrated up to $11$~MeV. In both cases, the force PC-F1 has been used. Panel (b): The neutron emission threshold for the same
range of Sn isotopes. Here the force PC-F1 is used.}
\label{fig04}
\end{figure}

A more careful study in this low-lying mode has shown that its determination is not easy for two main reasons. Firstly, the
experimental identification is very difficult on this area, since it often lies below the particle emission threshold,
where $(\gamma,n)$ reactions are no more possible.
The other reason is that the exact position of this soft mode has been found to be very sensitive to the basis truncation, i.e. to the choice of the
energy cut-off $E_{cut}^{ph}$. This can bee clearly seen in Fig.~\ref{fig05}, where we compare the E1 strength of $^{124}$Sn derived from
the discrete RPA approach with various energy cut-offs with results from continuum RPA. For the sake of the present discussion, a very small smearing
parameter $\Gamma$ has
been used. It is very interesting to see how the position of the pygmy mode moves to lower energies, as we increase the configuration space and also
that it approaches the position of continuum RPA, in which the entire configuration space is included.

\begin{figure}[b]
\centering
\includegraphics[width=240pt]{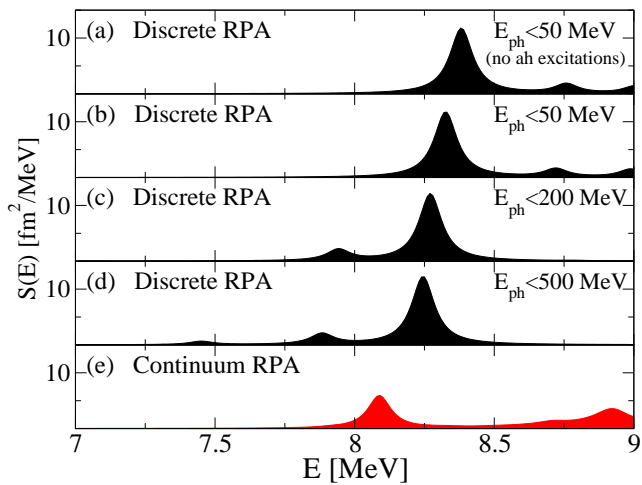}\newline\caption{(Color online) PDR strength using DRPA with differnt energy cut-off (panel a-d) and CRPA
(panel e).}
\label{fig05}%
\end{figure}

Therefore, it becomes evident that a proper treatment of the continuum seems to be very important in the calculation of the low-lying collective
phenomena. It has to be emphasized that till now in all the previous relativistic investigations this has not been possible. In Fig.~\ref{fig06} we go
further and show the details of the PDR in the isotope $^{132}$Sn, which has only recently been identified experimentally~\cite{AKF.05}.

The dashed line correspond to the discrete RPA, while the solid one shows the continuum RPA calculations. Although the experimental uncertainties are
still large in this neutron rich nucleus, we find a nice agreement with the Relativistic RPA model, in particular in the continuum approach.
In the right panel of Fig.~\ref{fig06}, the proton and neutron transition densities at the peak energy is used to show the pygmy character of this
mode, i.e. the oscillation of a relatively pure neutron skin against the isospin-saturated proton-neutron core.

\begin{figure}[!t]
\centering
\includegraphics[width=250pt]{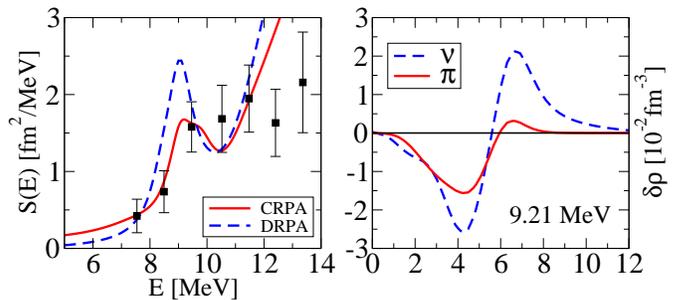}%
\newline\caption{Panel (a): The low-lying E1 strength distributions for $^{132}$Sn
using Discrete (blue dashed line) and Continuum RPA (red solid line). In both cases, the point coupling force PC-F1 has been applied. Panel (b): The
proton and neutron transition densities at the peak of the soft mode. The experimental data are taken from Ref.~\cite{AKF.05}.}
\label{fig06}
\end{figure}

\section{Conclusions}
\label{summary}

Starting from a point coupling Lagrangian, we have used the relativistic continuum QRPA approach to examine the E1 collective excitation spectra of
spherical open-shell nuclei. This non-spectral method has several advantages. The use of exact scattering wavefunctions with proper boundary
conditions, instead of expansions in a harmonic oscillator basis in the continuum allows for the simultaneous inclusion of the entire continuum and
the contributions of the Dirac sea. Furthermore, large sums over unbound states are avoided, which is extremely important in relativistic model, since
the unbound states in the Dirac sea are the root for computationally very expensive calculations.

In these investigations the ground state properties are calculated using a relativistic point coupling Lagrangian with the parameter sets PC-F1 and
DD-PC1. The RMF equations are solved in $r$-space self-consistently. For open-shell nuclei, the BCS model is applied to treat the pairing correlations
properly.

The residual particle-hole interaction used in the QRPA calculations is derived from the same parameter sets as the second derivative of the energy
functional with respect to the density. In this way no additional parameter is required for the study of the dynamical problem. One has current
conservation and an exact separation of the spurious modes and one is able to reproduce the collective properties, as for instance the multipole giant
resonances, in a fully self-consistent way.

The calculations are carried out by using two different point-coupling forces, namely the PC-F1 and DD-PC1. The interaction then includes the
basic zero range terms, rearrangement terms due to density dependance, the derivative terms which simulate the finite range of the nuclear
interaction, the various current-current terms and finally the Coulomb interaction between protons.

We have used the continuum QRPA approach to study collective properties, which are initiated by photo-absorption processes. We have shown that the
model performs well in describing both the giant dipole and the soft pygmy resonance. In particular, although there are not large differences in the
details between the continuum and the discrete RPA calculations as far as the GDR is concerned, there is still some systematic difference close to the
neutron separation threshold of stable nuclei and an even larger one for the extreme neutron-rich nuclei. New ($\gamma,\gamma^{\prime}$) experiments
on this low-lying area~\cite{TLA.10} are expected to be of paramount importance to the understanding of the real contribution of the exact
coupling to the continuum.

Finally, this approach accounts for nuclei far from the drip lines where no level in the continuum is occupied and thus the BCS models can be safely
applied. For nuclei up to the drip line, the relativistic Hartree-Bogoliubov approximation would be required. Investigations in this direction are in
progress.

\section*{Acknowledgments}

Helpful discussions with S. Goriely, G. Lalazissis, D. Pena Arteaga, T. Nik\ifmmode \check{s}\else \v{s}\fi{}i\ifmmode \acute{c}\else
\'{c}, and D. Vretenar are gratefully acknowledged. This work was financially supported by the FNRS (Belgium) and the Communaut\'e  fran\c{c}aise de
Belgique (Actions
de Recherche Concert\'ees) and by the DFG cluster of excellence \textquotedblleft Origin and
Structure of the Universe\textquotedblright\
(www.universe-cluster.de).


\end{document}